%% file: RiMa_BPMDS11.tex
\documentclass{llncs}
\usepackage{graphicx}
\usepackage{bigdelim}
\usepackage{multirow}
\usepackage{amssymb,amsmath}
\usepackage{amsfonts}
\usepackage{framed}
\usepackage{pifont}
\usepackage{wrapfig}
\usepackage{color}
\usepackage{subfigure}
\usepackage{pdfpages}
\usepackage{listings}
\usepackage{breakurl}
\usepackage{enumitem}

\setitemize{partopsep=0pt, parsep=0pt, itemsep=0pt}

\usepackage[utf8]{inputenc}

\RequirePackage[bookmarks,
                urlcolor=black,
                citecolor=black,
                linkcolor=black,
                pagecolor=black,
                colorlinks,
                hyperfigures]{hyperref}

\newcommand{\furl}[1]{\textit{\url{#1}}}

\title{{\sc IUPC:} Identification and Unification \\of Process Constraints}
\author{Juergen Mangler and Stefanie Rinderle-Ma}
\institute{%
  University of Vienna, Austria\\%
  Faculty of Computer Science\\%
  Workflow Systems and Technology Group\\%
  \{stefanie.rinderle-ma, juergen.mangler\}@univie.ac.at%
}

\begin{document}
  \maketitle              

  \begin{abstract}
    \input{abstract}
  \end{abstract}

  \input{introduction}
  \input{preliminaries}
  \input{structure}

  \input{properties}
  \input{evaluation}
  \input{related}
  \input{summary}

  \bibliographystyle{splncs03}
  \bibliography{compliance,cpee} 
\end{document}

%% file: abstract.tex
Business Process Compliance (BPC) has gained significant momentum in research
and practice during the last years.  Although many approaches address BPC, they
mostly assume the existence of some kind of unified base of process constraints
and focus on their verification over the business processes. However, it
remains unclear how such an integrated process constraint base can be built up,
even though this constitutes the essential prerequisite for all further
compliance checks. In addition, the heterogeneity of process constraints has
been neglected so far. Without identification and separation of process
constraints from domain rules as well as unification of process constraints,
the successful IT support of BPC will not be possible.  In this technical
report we introduce a unified representation framework that enables the
identification of process constraints from domain rules and their later
unification within a process constraint base. Separating process constraints
from domain rules can lead to significant reduction of compliance checking
effort. Unification enables consistency checks and optimizations as  well as
maintenance and evolution of the constraint base on the other side. 

%% file: introduction.tex
\section{Introduction}
\label{sec:introduction}

\begin{figure}
  \begin{center}
    \includegraphics[width=1.0\textwidth]{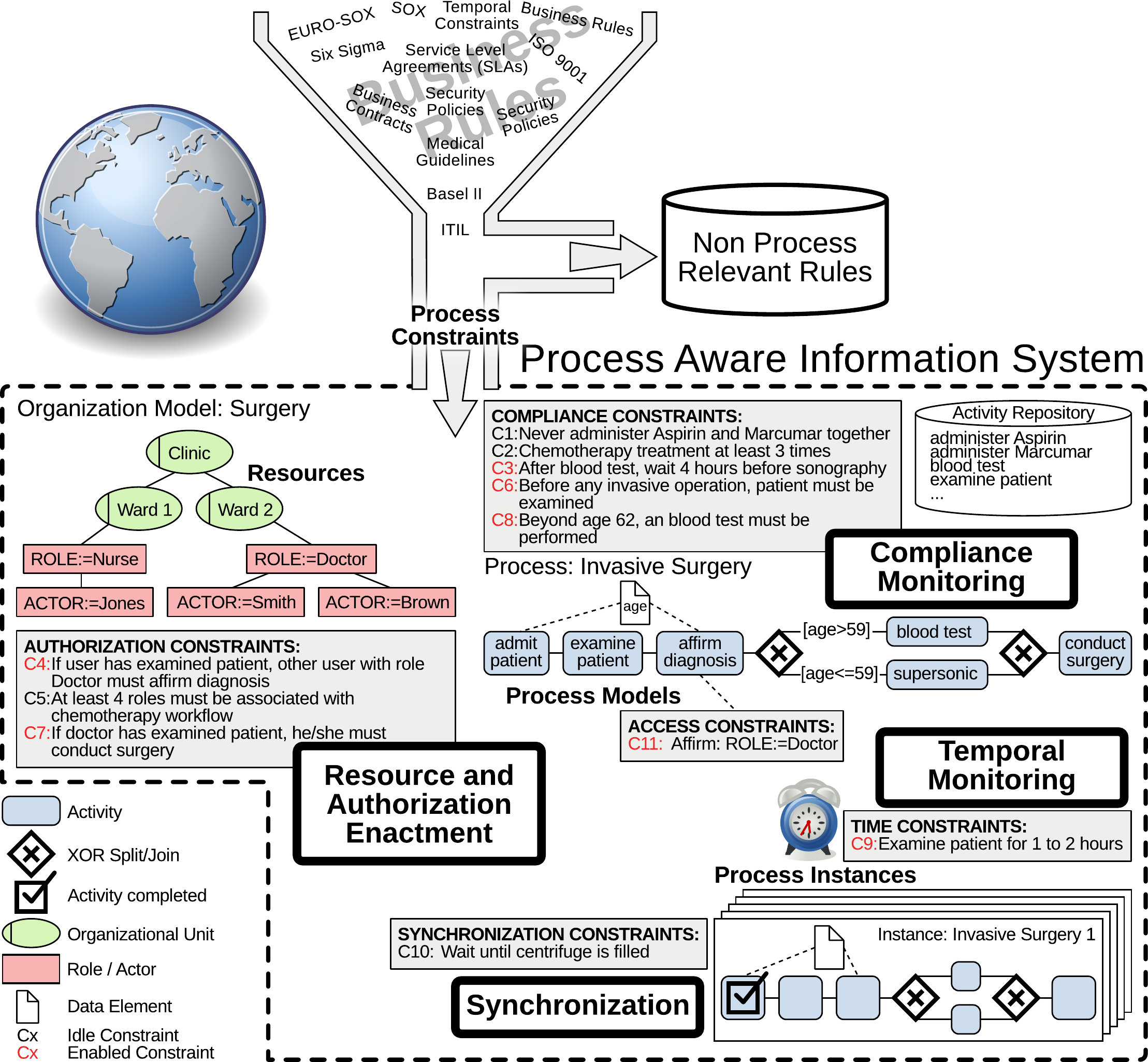}
    \caption{\label{fig:overall}Current Situation: ``Zoo'' of Process-Relevant Constraints}
  \end{center}
  \vspace{-23pt}
\end{figure}

Business Process Compliance (BPC) has gained significant momentum in research and practice during the last years. 
BPC requires that business processes comply with certain relevant rules, regulations, or norms. The rules can be derived from internal quality directives such as Six Sigma or ITIL. Examples for external rules comprise regulations by standards (e.g., ISO 001), regulations by authorizing bodies, or regulations based on contracts (business contracts) \cite{sadiq_modeling_2007}. 

Although many approaches address BPC \cite{namiri_pattern-based_2007,lu_compliance_2007,sadiq_modeling_2007,ly_design_2010,awad_visually_2011}, they mostly assume the existence of some kind of unified base of process constraints and focus on the verification of these constraints over the business processes. By {\sl process constraints} we refer to those rules in the domain of interest, that are associated with processes. As a general remark: the co-existence of processes and process constraints is common and desired (``Separate [Rules] From Processes, Not Contained In Them'', cf. \cite{business_rules_group_business_????}). 
However, it remains unclear how such an integrated process constraint base can be built up, even though this constitutes the essential prerequisite for all further compliance checks. 

In addition, the heterogeneity of process constraints has been mostly neglected so far. For a specific application scenario consider Fig. \ref{fig:overall} that displays example processes and rules from the medical domain: here, business
processes might be subject to medical guidelines or clinical
pathways on the one side \cite{lu_compliance_2007}
and also subject to authorization and privacy constraints on
the other side. Fig. \ref{fig:overall} also covers simple resource
synchronization as well as temporal rules. Assuming that all of these rules are process constraints, how can they be integrated within a unified representation?
In summary, building up a unified basis of process constraints out of all rules existing in a certain business context, basically poses the following requirements: 
 \begin{itemize}
 \item {\sc Requirement 1:} Process constraints should be identified and separated from the overall set of business rules.
 \item {\sc Requirement 2:} Process constraints should be unified based on a common representation in order to facilitate consistency checking and optimization.
 \end{itemize}

Requirement 1 is crucial in order to reduce BPC verification effort to those constraints that are associated with processes. This is particularly important since compliance verification for processes, which is mostly based on some kind of model checking technique, tends to be complex \cite{ly_design_2010}. The second requirement
demands for a unified representation that allows to fully match the heterogeneity of vastly different constraint types. It can then be used for filtering and optimization of process constraints. Furthermore, it also contributes to evolution and maintenance of the constraint base which is particularly important, since constraints are often subject to change and evolution \cite{ly_enabling_2009}. 

In this technical report, we present a definition for process constraints that enables their distinction from business rules that are not associated with any process. This definition bases on results from general business rules frameworks as well as analysis of existing BPC approaches and case studies. Further, a unified representation for process constraints is provided. It covers process perspectives such as control flow, data flow, time, and resources which can be subject to process constraints. The unified representation also enables the specification of behavior which is relevant for process constraints that require some sort of action (e.g., synchronization between process instances). 

In Sect. \ref{sec:preliminaries} we derive means to identify process constraints from domain rules. The IUPC unification representation is presented in Sect. \ref{sec:structure}. Sect. \ref{sec:properties} discusses the IUPC representation along different constraint properties in PAIS followed by an evaluation in Sect. \ref{sec:casestudies}. Sect. \ref{sec:related} compares related approaches and Sect. \ref{sec:summary} concludes with summary and outlook. 

%% file: preliminaries.tex
\section{Identification of Process Constraints} 
\label{sec:preliminaries}

This section addresses Requirement 1 as stated in the introduction: from all business rules relevant for ``business practice and guidance'' (cf. GUIDE \cite{business_rules_group_business_????}), those constraints that are relevant in the context of process execution should be filtered out. Reason is that the effort for checking constraints imposed over processes can be reduced to those which are actually associated with the process (structure). Note that verifying constraints over processes is often complex although different approaches offer optimization techniques \cite{ly_design_2010}. 

In order to be able to identify and separate process constraints from business rules, a definition of process constraints is required. In turn, a definition requires to know the specifics of process constraints in contrast to business rules. Intuitively, process constraints should be somehow associated with a process. In literature on business rules, association of business rules and real-world objects is expressed by so called anchor objects \cite{guide_business_rules_project_defining_2000}. In the example rule ``A car must have a registration number'', real-world object ``car'' constitutes the anchor object \cite{guide_business_rules_project_defining_2000}. More precisely, within the generally accepted  ``If ... Then'' rule structure, the anchor object would be contained within the if part, triggering execution of the rule. The question is now which anchor objects are suitable to distinguish a process constraint from a general business rule. 

As process literature study shows, all formalisms on process constraints (implicitly) contain a process-specific anchor object, i.e., a structural pattern contained within the process of interest which the process constraint refers to. A structural pattern contains at least one activity either executed within a process stored within the process activity repository. An example for the first case is constraint C3 depicted in Fig. \ref{fig:overall}, referring to process activity {\tt blood test} executed within the treatment process. We denote such process constraints as {\sl enabled}. By contrast process constraint C1 for example does not refer to any process activity currently executed within the treatment process. In turn, it refers to processs activities {\tt administer Aspirin} and {\tt administer Marcumar}, stored within the associated process activity repository. Hence, any time one of these two activities will be added to some process, C1 will become enabled and thus is to be considered as process constraint. Until then, we denote C1 as {\sl idle}.

Specifically, we use the general term structural pattern, since it might refer to a set of process activities as well as to control flow patterns \cite{van_der_aalst_workflow_2003}, e.g., a sequence or a parallel branching. 

\begin{definition}[Process Constraint]
\label{def:processconstraint} 
Let ${\cal P}$ be a set of all process types of consideration. Let further ${\cal A}$ := ${\cal A_P}$ $\dot{\cup}$ ${\cal A_R}$ be the set of all process activities within the domain where ${\cal A_P}$ denotes the set of process activities executed within a process P $\in$ ${\cal P}$ and ${\cal A_R}$ denotes the set of process activities stored within the associated process repository.\footnote{For definition purposes we assume disjoint sets here. However, the definition can be easily adapted when used and idle process activities are contained in both sets.} Finally, let ${\cal R}$ be the set of all domain rules. 
Then we denote a rule r $\in$ ${\cal R}$ as process constraint if it contains a structural pattern SP$_c$ as anchor object with SP$_c$ is structural pattern over ${\cal A_P}$ or SP$_c$ $\subseteq$ ${\cal A_R}$. 
\end{definition}

In the SeaFlows project \cite{ly_design_2010} process constraints are tied to one or more process activities and consist of an {\sl antecedent} and a {\sl consequence} part. More precisely, based on a triggering antecedent pattern, a consequence pattern must hold to fulfill the constraint. This directly corresponds to the general notion of anchor object, since the structural pattern contained within the antecedent pattern triggers the constraint (if part). Process constraints as defined in DECLARE also refer to process activities and the semantics for relations between activities are based on LTL (Linear Temporal Logic) \cite{pesic_constraint-based_2007}. BPMN-Q offers compliance patterns containing structural patterns within the triggering part of the constraint \cite{awad_visually_2011}. 
Consequently, we can state that a business rule is a process constraint if its anchor object contains a structural pattern within a process.

%% file: structure.tex
\section{Process Constraint Unification}
\label{sec:structure}

As we can now identify process constraints, the second challenge is to unify the partly strongly varying process constraints. Hence, in this section we present the IUPC representation for unifying and structuring process constraints in to subsequently support process constraint optimization, maintenance, and evolution. 

\subsection{Expressing Process Context by Linkage}
\label{sub:process context}

The following definition of {\sl Linkage} integrates the context
$\texttt{Context}_c$ of a process constraint $c$ together with its structural
pattern $\texttt{SP}_c$ and a trigger position $\texttt{TP}_c$ (cf. Fig. \ref{fig:structure}).

\begin{definition}[Linkage]
\label{def:linkage} 
Let ${\cal P}$ be a set of all process types of consideration and ${\cal I}_P$ be the set of process instances running according to a process type $P~\in~{\cal P}$. A process type P $\in$ ${\cal P}$ is described by a process schema $S_P$ := ($A_P$, $E_P$, $D_P$)\footnote{In order to stay meta-model independent, we assume a simple generic representation for process schemas that can be adopted by any (imperative) process meta model.} where $A_P$ denotes the set of activities, $E_P$ the set of control/data edges, and $D_P$ the set of data elements $S_P$ consists of. Then the linkage ${Linkage}_c$ to a process type $P$ is defined as follows:
  \begin{align*}
     \texttt{Linkage}_c           & \subseteq \texttt{Context}_c \times \texttt{SP}_c \times \texttt{TP}_c &\textit{where} \\
     \texttt{Context}_c           & \subseteq {\cal P} \times {\cal I}_P & \wedge \\
     \exists\texttt{SP}_c & & \wedge \\
     \texttt{TP}_c   & \in \emptyset, \texttt{before}(a_{n,P}), \texttt{after}(a_{n,P}) 
  \end{align*}
\end{definition}

In Def. \ref{def:linkage}, $\texttt{Context}_c$ describes in which processes or
process instances a constraint may occur. Possibilities include single instances (e.g. SLAs for services that have been dynamically selected), all instances of a process (attribution of resources to tasks) or even instances in multiple processes (when a resource is used in multiple processes, and synchronization has to take place). Structural patterns $\texttt{SP}_c$ express the association between process constraint and process. In connection with {\sl context}, it defines for which parts of process types and process instances the corresponding process constraint is to be enforced or verified. Structural patterns may not only spawn single activities but also several activities connected through complex control decisions. Finally, the trigger position $\texttt{TP}_c$ is relevant for synchronization constraints, since synchronization constraints are constraints that not only set out certain conditions on process execution, but enforce an action. In this case the trigger position specifies whether the action should take place before the affected activity is started or after. The Trigger Position is solely present for run-time (behavioral) constraints. As a structural pattern may not only spawn a single activity but also several activities connected through complex control decisions, it is necessary to determine when exactly a constraint has to fire. This is the equivalent of describing the condition under which an event occurs, in an Event Condition Action (ECA) rule. The trigger position itself is simple: before or after an activity occurs. Of course multiple before / after positions can be specified.

Both, $\texttt{SP}_C$ and $\texttt{TP}_c$ are grouped as \textit{Connection} in Fig. \ref{fig:structure} to express their tight integration. A $\texttt{TP}_c$ can not exists without a set of activities matched by a structural pattern $\texttt{SP}_c$. Apart from $\texttt{TP}_c$, linkage information can be statically matched against processes, thus it is possible to determine \textit{enabled} and \textit{idle} constraints (as described before). 

From now on we will denote a linkage $\texttt{Linkage}_c$ in the compact form:
\[
  \texttt{Linkage}_{c}: (({\cal P},{\cal I}_P), \texttt{SP}_c, \texttt{TP}_c)
\]

Consider compliance constraint C6 depicted in Fig. \ref{fig:overall}: ``{\tt C6: Before any invasive operation, patient must be examined}''.  

The antecedent pattern ``{\tt existence of activity invasive operation}'' within a process
triggers the check whether this activity is preceded by an activity ``{\tt patient
examination}''. The linkage for this compliance constraints turns out as
\[
  \texttt{Linkage}_{C6}: ((\texttt{Invasive Surgery}, \texttt{ALL}), \texttt{SP}_{C6}, \emptyset) 
\]  
\noindent{}with 
\begin{align*}
  \texttt{SP}_{C6}: &\ \exists a_1\ Is(a_1, \texttt{examine patient}) & \wedge \\
                    &\ \exists a_2\ Is(a_2, \texttt{conduct surgery}) & \wedge \\
                    &\ a_1{\cal A}^*a_2
\end{align*}
\noindent{}where 
\[
  {\cal A}^*: \textit{arbitrary activities between}\ a_1\ \textit{and}\ a_2
\]  

\subsection{Integrating Data, Time, and Resources}
\label{sub:data}

Existing approaches mostly deal with control flow constraints, i.e.,
constraints that are only referring to structural patterns within a process
\cite{awad_visually_2011}. The only approaches that have addressed data flow
aspects within process constraints are SeaFlows \cite{knuplesch_enabling_2010}
and BPMN-Q \cite{awad_visually_2011}. In accordance to these approaches, the
data flow perspective within a process constraint can be represented as
condition on the structural pattern it refers to. Consider constraint C8 from
Fig. \ref{fig:overall}: ``{\tt C8: Beyond age 62, a blood test must be
performed}''.

We can see that the data flow condition ``{\tt Beyond age 62}'' imposes a restriction on the structural pattern of the constraint (``{\tt blood test}'').  
However, control and data flow are only two perspectives of a process. As case studies show, process constraints might also refer to conditions on time and resources, e.g., constraints C3, C4, C5, C7, C9, and C11 within the example depicted in Fig. \ref{fig:overall}. Hence, constraint unification requires the specification of data, time, and resource conditions on top of the linkage part of the process constraint. 

\begin{definition}[Condition]
\label{def:condition}
Let $c$ be a process constraint with linkage $\texttt{Linkage}_c$ referring to a process type P described by process schema on linkage $S_P$ := ($A_P$, $E_P$, $D_P$). The condition of $c$ imposed on $\texttt{Linkage}_c$ is defined as 
\begin{align*}
     \texttt{Condition}_c & \subseteq \texttt{expr($D_P$)} \times \texttt{expr(time$_c$)} \times \texttt{expr(resource$_c$})
\end{align*}
where 
\begin{itemize}
  \item expr($D_P$) denotes a logical expression over the data elements of P
  \item expr(time$_c$) denotes a temporal expression and
  \item expr(resource$_c$) denotes a logical expression over the resources associated with P (typically modeled within a organizational or resource model)
\end{itemize}
\end{definition}

It is important to mention that a condition cannot exist without referring to a linkage. Thus, a condition describes, under the premise of a given linkage, either (a) a combination of data, temporal or resource conditions that have to be present or (b) a combination of data, temporal or resource conditions that have to be present in order for a given behavior to be apply.

For constraint C6, for example, no specific condition is imposed on the linkage part. Hence:
\[
  \texttt{Condition}_{C6}: \emptyset 
\]  

When considering constraint C3 in Fig. \ref{fig:overall}, things get more interesting, since ``{\tt After a blood test wait 4 hours before sonography}'' obviously imposes a time condition on the linkage part. Less obviously, an additional condition is imposed on the data flow, since the 4 hours time frame between blood test and sonography are only required for the same patient, formally: 
\[
  \texttt{Linkage}_{C3} : (\texttt{Invasive Surgery}, \texttt{ALL}), \texttt{SP}_{C3}, \emptyset) 
\]  
with 
\begin{align*}
  \texttt{SP}_{C3}: &\ \exists a_1\ Is(a_1, \texttt{blood test}) & \wedge \\
                    &\ \exists a_2\ Is(a_2, \texttt{sonography}) & \wedge \\
                    &\ a_1{\cal A}^*a_2
\end{align*}
and 
\begin{align*}
  \texttt{Condition}_{C3}: &\ \textit{patient}(a_1) = \textit{patient}(a_2) & \wedge \\
                           &\ \textit{min\_time\_between}(a_1,a_2,4h)
\end{align*}

Similar considerations can be made for data (``C8: {\tt beyond age 62}'') and resources
(C7: {\tt examination by same user as surgery}) which can be also represented by a
condition part imposed on the linkage of a constraint. 

\subsection{Expressing Behavior within Process Constraints}
\label{sub:behavior}

The last missing piece is, given a certain linkage and condition, to allow for a certain assignment or behavior. Assignment becomes necessary for access constraints such as C11. Behavior specifies for example the action part of a synchronization constraint such as C10. Hence, we complete the unified constraint representation by a {\sl Behavior} part.

\begin{definition}[Behavior]
\label{def:behavior}
Let $c$ again be a process constraint. For a given $\texttt{Linkage}_c$ and $\textit{Condition}_c$, a behavior can be either empty, the attribution of resource or time information or instructions specific for process execution (e.g. exceptions).
\end{definition}

\subsection{Summary: Linking, Condition, Behavior}
\label{sub:behavior}

\begin{figure}
  \begin{center}
    \includegraphics[width=0.8\textwidth]{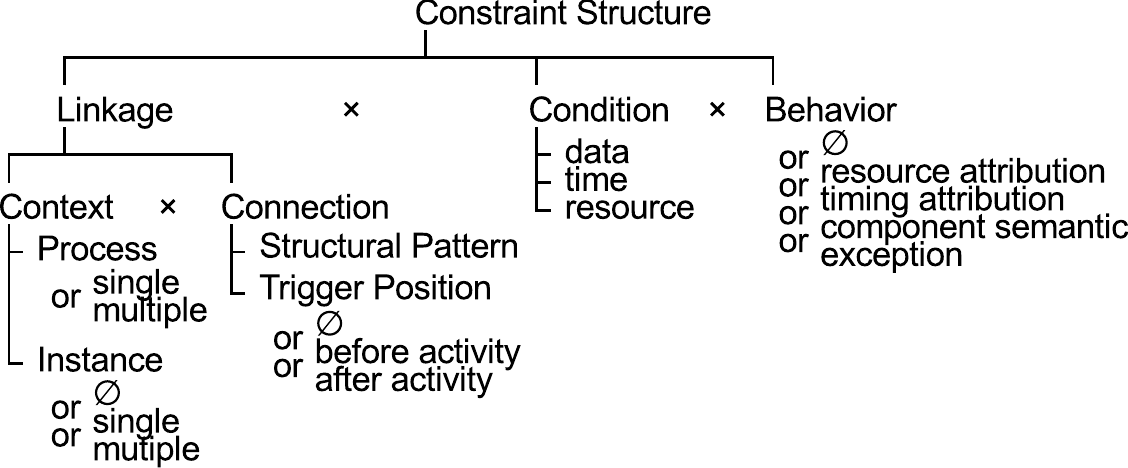}
    \caption{\label{fig:structure}Constraint Structure}
  \end{center}
  \vspace{-23pt}
\end{figure}

The overall representation for process constraints consisting of linkage, condition, and behavior is summarized in Fig. \ref{fig:structure}. In Sect. \ref{sec:casestudies} we will evaluate the unified representation against constraints as shown Fig \ref{fig:overall}. 

%% file: properties.tex
\section{Discussing Unified Representation along Constraint Properties in Process-Aware Information Systems}
\label{sec:properties}

\begin{figure}
  \begin{center}
    \includegraphics[width=0.6\textwidth]{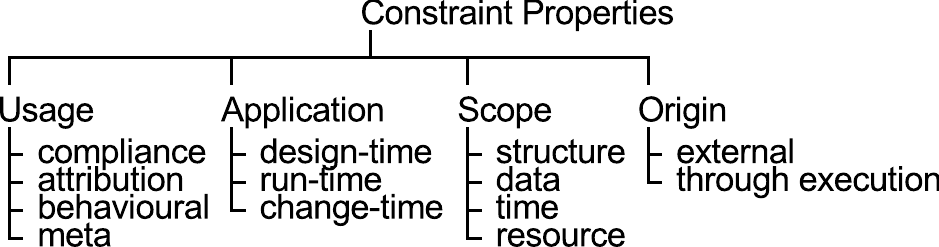}
    \caption{\label{fig:properties}Process Constraint Properties}
  \end{center}
  \vspace{-23pt}
\end{figure}

In the previous sections, a unified representation for process constraints has been presented. In this section, we discuss its usage within Process-Aware Information Systems (PAIS) along different constraint properties as set out in Fig. \ref{fig:properties}. 

\noindent{\bf Usage:} Process constraints can be used to verify the compliance of business processes with relevant regulations and constraints. This is useful for process design and evolution. Constraints can also be used to alter or affect the behavior of processes (\textit{behavioral}). Affecting the behavior of processes includes the attribution of process activities or, more generic, the attribution of structural patterns. Examples include resources allocation of process activities.  

Additionally, constraints can be used to specify \textit{meta} constraints. Meta constraints are intended to check the consistency of other constraints, for example, constraint C5 in Fig.  \ref{fig:overall}. Another example is a meta constraint specifying that for each process activity referring to a resource {\tt centrifuge} synchronization constraint C10 in Fig. \ref{fig:overall} is assigned to. 

\noindent{\bf Application:} Application deals with the question at which phase of a process life cycle constraints may be enforced. Basically, at \textit{design-time} constraints are checked to verify that a process complies to certain criteria (compliance) constraints. All constraints that are checked at design-time are \textit{compliance} constraints or \textit{meta} constraints.

All \textit{behavioral} constraints, and some \textit{compliance} constraints are checked at \textit{run-time}. Examples include the checking for data value constraints, or the attribution of structural patterns (the attributes are used at run-time, though can be checked by meta constraints at design-time). Compliance constraints may affect run-time under certain circumstances. E.g. although a process may not conform to a specific constraint at design-time, it may do so at run-time, because only a certain execution path violates the constraint. Hence corresponding process instances are to be monitored at run-time \cite{ly_design_2010}. 


\noindent{\bf Scope:} The four perspectives of the constraint scope have been discussed in Sect. \ref{sec:preliminaries}, capturing which kind of information constraints a PAIS can contain. The most important scope perspective is \textit{structure} as in all constraints structural information has to be present (either explicit or implicit through connections of information to structure), as otherwise they would not be connected to processes and could thus not be enforced in PAIS. Note that if a constraint holds on only structural information, it is always a \textit{compliance} constraint.

\textit{Data} is also an integral part of a process, that is tightly connected to structure. At design-time, it is possible to check if data types and data flow conform to a certain schema. Further it can be checked whether certain data values will lead to compliance violations are run-time. If, for example, a treatment process states that a lab test is to be performed for all patients beyond $65$ years and the corresponding medical guideline states that the lab test is mandatory for patients beyond $62$ year, it can be already checked at design-time, that for patients between $62$ and $64$ there will be a violation of the corresponding constraint at run-time.  

\textit{Resource} and \textit{time} are attributes that are typically connected to structure (or data). Their purpose is to describe information that aids the execution of a process.  {\sl Resource-aware} and {\sl time-aware} constraints in PAIS are typically checked or enforced at run-time including:

\begin{itemize}

\item Resource assignment to process activities, typically specified based on access constraints (e.g., constraint C11 in Fig. \ref{fig:overall}). Based on role assignment, process activities are offered to authorized actors in their work lists at run-time (e.g. by a RBAC component). 

\item On top of access constraints, authorization constraints can be specified such as dynamic separation of duties. Dynamic authorization constraints are verified during run-time.  

\item Assign temporal information to activities (e.g. the normal duration of a certain activity is 2 hours with a standard deviation of 10 minutes).
  

\end{itemize}

\noindent{\bf Origin:} When do constraints become available for application to a process? All constraints become available through not specified \textbf{external} resources, either at design time, run-time or change time. They are identified and structured by constraint designers and then made available through a constraint-base, and have to be enacted from this point on. One exception is constituted by SLAs for dynamic service selection. These constraints become available \textit{through execution} of a process instance, and vanish after the instance finishes. 

%% file: evaluation.tex
\section{Evaluation} 
\label{sec:casestudies}

The feasibility of the unified representation is evaluated by means of the overall 16 process constraints depicted in Fig. \ref{fig:overall} and along the constraint properties described in Section \ref{sec:properties}. For this we classify the 16 process constraint into different constraint types that embody a combination of properties as represented by the unified representation. 

\begin{table}
  \vspace{-20pt}
  \begin{center}
    \includegraphics[width=1.0\textwidth]{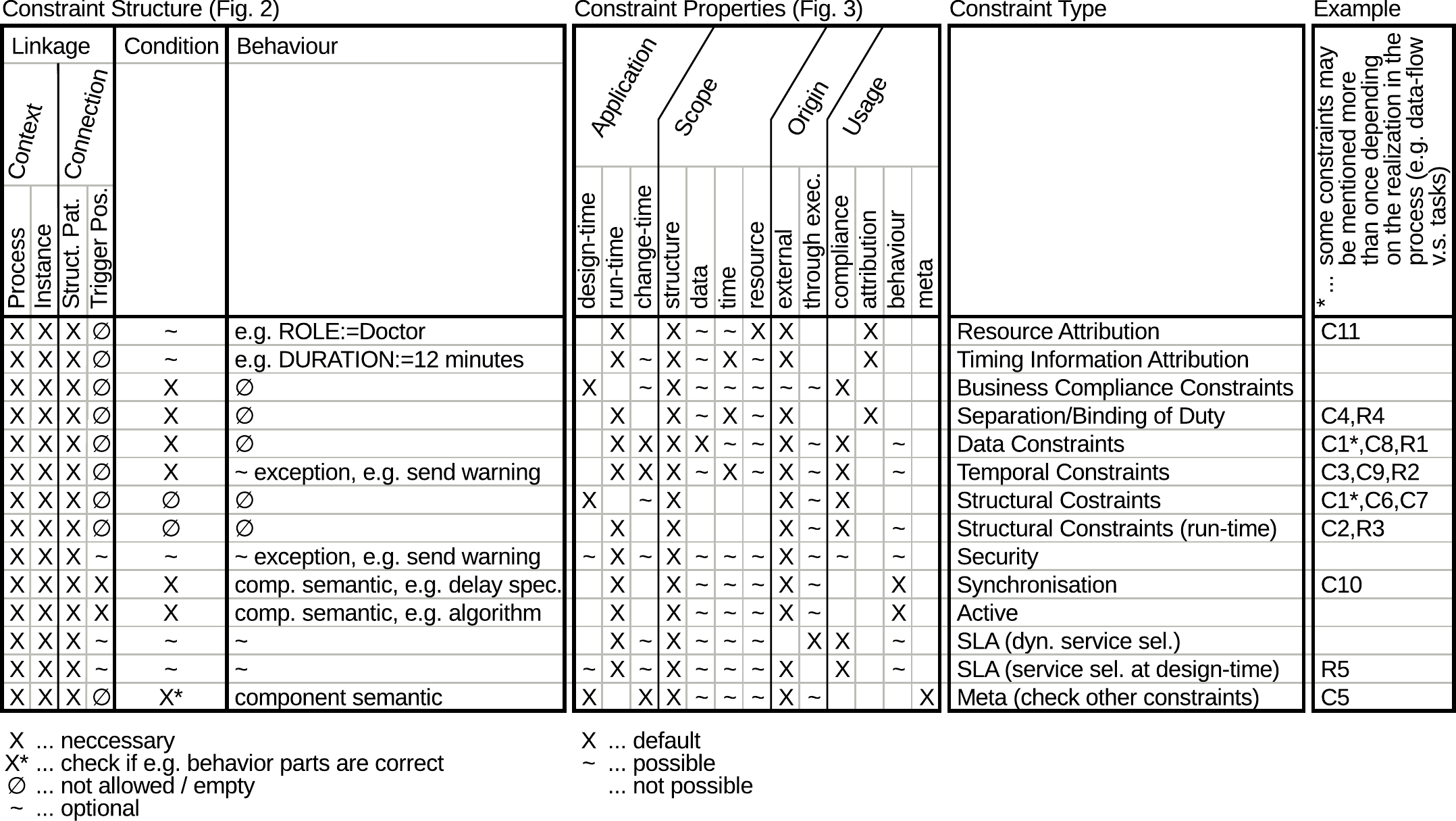}
    \caption{\label{tab:types}Constraint Types}
  \end{center}
  \vspace{-23pt}
\end{table}

As depicted in Tab. \ref{tab:types} we identified 14 different process constraint types. The first two constraint types deal with \textit{Resource Attribution} (e.g., roles, actors, nodes) and \textit{Timing Information Attribution} (e.g., minimal, maximal, average duration) to process structure. All this information is necessary at run-time, for a temporal monitor as a basis to verify if a process behaves correctly, or an RBAC system to select actors in a worklist. The only difference between these two is, that during change-time timing information may have to be adapted to account for the duration of changes. E.g. constraint C11 in Fig. \ref{fig:overall} can represented as follows:
\begin{align*}
  \texttt{Linkage}_{C11}:   &\ ((\texttt{Invasive Surgery}, \texttt{ALL}), \texttt{SP}_{C11}, \emptyset) & \textit{with} \\
  \texttt{SP}_{C11}:        &\ \exists a_1\ Is(a_1, \texttt{affirm diagnosis}) \\
  \texttt{Condition}_{C11}: &\ \emptyset \\
  \texttt{Behavior}_{C11}:  &\ \{\textit{ROLE}:=\textit{Doctor}\}
\end{align*}

For attribution $\texttt{TP}_c$ is always empty, as the matched structural pattern implies, that the attribution is available all the time.

The third constraint type describes \textit{Business Compliance Constraints} in general. We decided to include this very generic constraint type (which should not be confused with several of the other constraint types) to show some specific characteristics we derived by studying constraint sources as mentioned in Section \ref{sec:preliminaries}. Business compliance constraints, are exclusively compliance constraints, they never carry a behavioral part, they most of the time verify a certain structure of the process, with a possibility of checking also data, resource and temporal aspects. Therefore the design-time \textit{Structural constraints} also in this list are \textit{Business Compliance Constraints}, as well as some \textit{Temporal Constraints} or \textit{Data constraints} (whenever only design-time aspects are covered). \textit{Separation / Binding of Duty}, \textit{Resource Attribution}, and \textit{Timing Information Attribution} are the exception to the rule. They may be very well applied at design time, just only the impact can only be seen at run-time.

%% file: related.tex
\section{Related Work}
\label{sec:related}

Many approaches for checking BPC either at design or run time exist, e.g., \cite{lu_compliance_2007,sadiq_modeling_2007,governatori_compliance_2006}).
As argued before, identification and unification of process constraints has been outside the scope of these approaches so far. Validation of the IUPC representation compared to selected existing approaches is presented in Table \ref{tab:comp}. 

\begin{table}
\begin{center}
\begin{scriptsize}
\begin{tabular}{p{2.4cm}p{1.8cm}p{2.2cm}p{2.2cm}p{2.4cm}} \\\hline
{\bf Representation} & SeaFlows \cite{ly_design_2010} & DECLARE \cite{pesic_constraint-based_2007} & BPMN-Q \cite{awad_visually_2011} & IUPC\\\hline
{\sl Linkage} & activity set & control flow pat. & control flow pat. & control flow pat. \\
              &              &                   & global scope      & process context \\
              &              &                   &                   & trigger position \\\hline
{\sl Condition} & data & -- & data & data, time, resource \\\hline
{\sl Behavior} & -- & -- & -- & $\checkmark$ \\\hline
\end{tabular}
\caption{\label{tab:comp} Comparison of Existing Approaches}
\end{scriptsize}
\end{center}
\vspace{-23pt}
\end{table}

The above mentioned approaches try to ensure BPC either a-priori at design time or by detecting inconsistencies at run-time. In addition, there are also a-posteriori approaches that offer techniques to analyze process logs (i.e., data of already executed and finished processes) with respect to certain properties such as adhering to compliance constraints \cite{aalst_process_2005}. In these approaches, different aspects logged during process execution can be checked, e.g., separation of duties. Doing so the a-posteriori approach reflects the need for unified compliance checking, not only a-posteriori, but also a-priori. 
Other approaches use constraints to synchronize between business processes of different type (e.g., between a chemotherapy and a radiation process) \cite{heinlein_workflow_2001}. Based on our unification approach, synchronization constraints can be unified and managed in combination with the other constraints in PAIS.

%% file: summary.tex
\section{Summary and Outlook}
\label{sec:summary}

This technical introduced the IUPC representation for identification and
unification of process constraints. By separating process constraints from
domain rules, effort for compliance verification can be significantly reduced.
Offering a unified representation enables consistency checks and optimizations
of the constraint base. Furthermore, unification and structuring of constraints
in PAIS enables the development of an integrated checking component instead of
isolated and distributed checking components as present nowadays.

In upcoming publications we provide (on the basis of this classification)
extensive evaluation based on case studies from different domains, e.g., health
care and financial sector. We furthermore work towards an implementation of the
unified and structured IUPC representation on top of our adaptive cloud process
execution engine CPEE \cite{sturmer_building_????}. Particular focus will be
put on maintenance of the constraint base and an integrated verification
component for the engine.